\documentclass[twocolumn]{aastex62}

\usepackage{amssymb,amsmath}

\usepackage{natbib}
\usepackage{hyperref}
\usepackage{array}
\usepackage{afterpage}
\usepackage{graphicx}

\newcommand{\Teff}{T_{\mathrm{eff}}}
\newcommand{\logg}{\log{g}}

\newcommand{\logRHK}{\log{R'_{\mathrm{HK}}}}
\newcommand{\SHK}{$S_{\mathrm{HK}}$}

\newcommand{\DeltaMG}{\Delta G}

\newcolumntype{C}[1]{>{\centering\let\newline\\\arraybackslash\hspace{0pt}}m{#1}}

\begin{document}

\title{Properties of F Stars with Stable Radial Velocity Timeseries: A Useful Metric for Selecting Low-jitter F Stars}

\author{Jacob K. Luhn}
\affiliation{Department of Astronomy, The Pennsylvania State University,  525 Davey Lab, University Park, PA 16802, USA}
\affiliation{Center for Exoplanets and Habitable Worlds, 525 Davey Lab, The Pennsylvania State University, University Park, PA, 16802, USA}
\affiliation{NSF Graduate Research Fellow}

\author{Jason T. Wright}
\affiliation{Department of Astronomy, The Pennsylvania State University,  525 Davey Lab, University Park, PA 16802, USA}
\affiliation{Center for Exoplanets and Habitable Worlds, 525 Davey Lab, The Pennsylvania State University, University Park, PA, 16802, USA}

\author{Howard Isaacson}
\affiliation{Astronomy Department, University of California, Berkeley, CA, USA}


\email{jluhn@psu.edu}

\begin{abstract}
In a companion paper, we have conducted an in-depth analysis of radial velocity jitter of over 600 stars, examining the astrophysical origins including stellar granulation, oscillation, and magnetic activity. In this paper, we highlight a subsample of those stars, specifically the main sequence and ``retired" F stars -- which we refer to as ``MSRF'' stars -- that show low levels of RV jitter ($<10$ m/s). We describe the observational signatures of these stars that allow them to be identified in radial velocity planet programs, for instance those performing followup of transiting planets discovered by TESS. We introduce a ``jitter metric" that combines the two competing effects of RV jitter with age: activity and convection. Using thresholds in the jitter metric, we can select both ``complete" and ``pure" samples of low jitter F stars. We also provide recipes for identifying these stars using only Gaia colors and magnitudes. Finally, we describe a region in the Gaia color-magnitude diagram where low jitter F stars are most highly concentrated. By fitting a 9th order polynomial to the Gaia main sequence, we use the height above the main sequence as a proxy for evolution, allowing for a crude selection of low jitter MSRF stars when activity measurements are otherwise unavailable.
\end{abstract}

\section{Introduction}\label{sec:introduction}
Intrinsic stellar phenomena can induce spurious radial velocity signatures and continue to be a large hurdle in the search for low mass exoplanets. These velocity variations, termed ``jitter" can come from a number of different sources. Stellar magnetic activity can produce star spots and flares, which suppress the local convective blueshift on the surface of the star, introducing variations on timescales of the rotation of the star. In addition, stars with activity cycles (similar to the 11 year solar cycle\footnote{In truth, the solar cycle is a 22 year cycle, whereby the magnetic poles switch every 11 years. However, in general the ``solar cycle" refers to the 11 year cycle of sun spot variations.}) can see RV variations that correlate with global magnetic field strength over long timescales. Convective motions in the star can also induce RV variations. On the surface of the star, convective motions manifest as granular regions of hot uprising material that cools and falls via a network of inter-granular lanes. While the mean effect of granulation is to produce the so-called ``net convective blueshift", the stochastic nature of the granulation network leads to short-term variations in the measured radial velocity. Further, the convective motions deep in the interior of the star drive pressure waves that resonate throughout the star in stellar oscillations. The deformation of the star from these oscillations can add additional RV noise.

Many studies have investigated these various sources of RV jitter. Early work focused on the relation between a star's magnetic activity and its radial velocity jitter \citep[e.g.]{Campbell1988,Saar1998, Santos2000,Wright2005, Isaacson2010}, finding that more active stars tend to exhibit higher RV jitter. Among these results was the fact that active F stars exhibit larger RV jitter than the active G and K stars \citep{Saar1997,Wright2005,Isaacson2010}. Combined with that is fact that F stars are near the Kraft break, $\Teff \sim 6200$~K \citep{Kraft1967}, where dramatically increased rotational velocities lead to increased rotational Doppler broadening, which in turn leads to less precise RV measurements. For these two reasons, RV surveys have largely avoided F stars due to their expected high levels of jitter. 

More recent work has involved the effects of convective motions such as granulation and oscillations \citep{Dumusque2011a,Bastien2014}, noting that evolved stars increase in jitter as they continue to evolve. Recently, \citet{Luhn2020}, L20 hereafter, showed a clear transition exists between the activity-dominated regime and the convection-dominated regime as stars evolve across the main sequence and onto the subgiant branch. In particular, they describe the jitter floor for stars from zero-age main sequence (ZAMS) through the subgiant phase at the bottom of the red giant branch. That is, stars across a wide range of masses start their main sequence lifetime as an active, jittery star. From there, they spin down, and decrease in both activity and jitter. As they continue to become more and more magnetically quiet, they are simultaneously changing their structure and the effect of granulation becomes increasingly important as the typical size of a granular region increases and oscillation power increases.

In addition to describing the jitter floor, L20 identified the jitter minimum for a wide range of masses, which is where contribution from all major components of RV jitter (activity, granulation, and oscillation) is minimized. Naturally this occurs at the transition from activity-dominated on the main sequence to convection-dominated among giants and subgiants. Furthermore, they showed that the jitter minimum occurs at later evolutionary stages (lower $\logg$) for higher mass stars, but that the stars in the jitter minimum are able to reach a level of RV jitter ($\lesssim 5$~m/s) similar to that of the lower mass G and K dwarfs (and their slightly evolved subgiant counterparts). This has particular implications for the F stars, which as described above have been largely avoided in RV surveys. 

To search for planets around stars of intermediate mass, surveys like the ``Retired" A Star survey \citep{Johnson2006} targeted subgiant stars, whose cooler temperatures and lower rotational velocities allow for precise Doppler work. These surveys have found that the population of planets around these stars is different from that around the lower mass G and K dwarfs in that there are typically more Jupiter-sized planets at further separations (1-2 au) \citep{Johnson2011}. Specifically, there is an apparent dearth of planets inside 1 au. As the name suggests, these are usually the evolved counterparts of main sequence A stars ($>1.5$~M$_{\odot}$). The F stars (1.1 - 1.4 M$_{\odot}$) are therefore at the boundary of the surveys that target ``sun-like" stars and the surveys that target intermediate mass stars and represent an important sample for bridging the gap between the observed differences between the populations of planets around these two groups of stars. 

In this paper, we focus on specifically these stars, the late F and early G stars, highlighting some of the observational properties of the RV stable stars. We provide tips on how to select the F-stars that are likely to have the lowest jitter value and will therefore be best for discovering planets. \autoref{sec:data} describes the data and selection of the `F' star sample used in this work. \autoref{sec:jitter_metric} describes the jitter metric we use in this work to distinguish between low and high jitter `F' stars. In \autoref{sec:low_jitter_f_stars} we use the jitter metric to provide thresholds for distinguishing between the low and high jitter stars. We then apply the same thresholds to a new sample of stars defined using readily-available Gaia data to try to reproduce the sample of `F' stars in \autoref{sec:restricted}. Finally, in \autoref{sec:gaia_only} we restrict ourselves even further to using only Gaia data (removing the need for activity measurements) to select low jitter `F' stars. We present conclusions and a summary in \autoref{sec:summary}

\section{Data Selection and Stellar Properties}\label{sec:data}
For this analysis, we use the data from L20, who performed an in-depth analysis of RV jitter on a star-by-star basis. These were stars with $>10$ RV measurements from the iodine RV work at Keck (most of which are also published as RVs in \citet{Butler2017}, all available at the Keck Observatory Archive).These values of RV jitter provide the best sample of RV jitter for a large sample of stars due to the treatment of known and suspected planets, long term linear trends, as well as careful analysis for stellar activity cycles and activity-correlated RV's. 

We adopt the spectroscopically-derived parameters of \citet{Brewer2017} because they have shown good agreement with stellar parameters (especially $\logg$) derived asteroseismically \citep{Brewer2015}. 

\subsection{Selecting `F' stars}\label{sec:f_stars}
\begin{figure*}
\centering
\includegraphics[width=0.415\paperwidth]{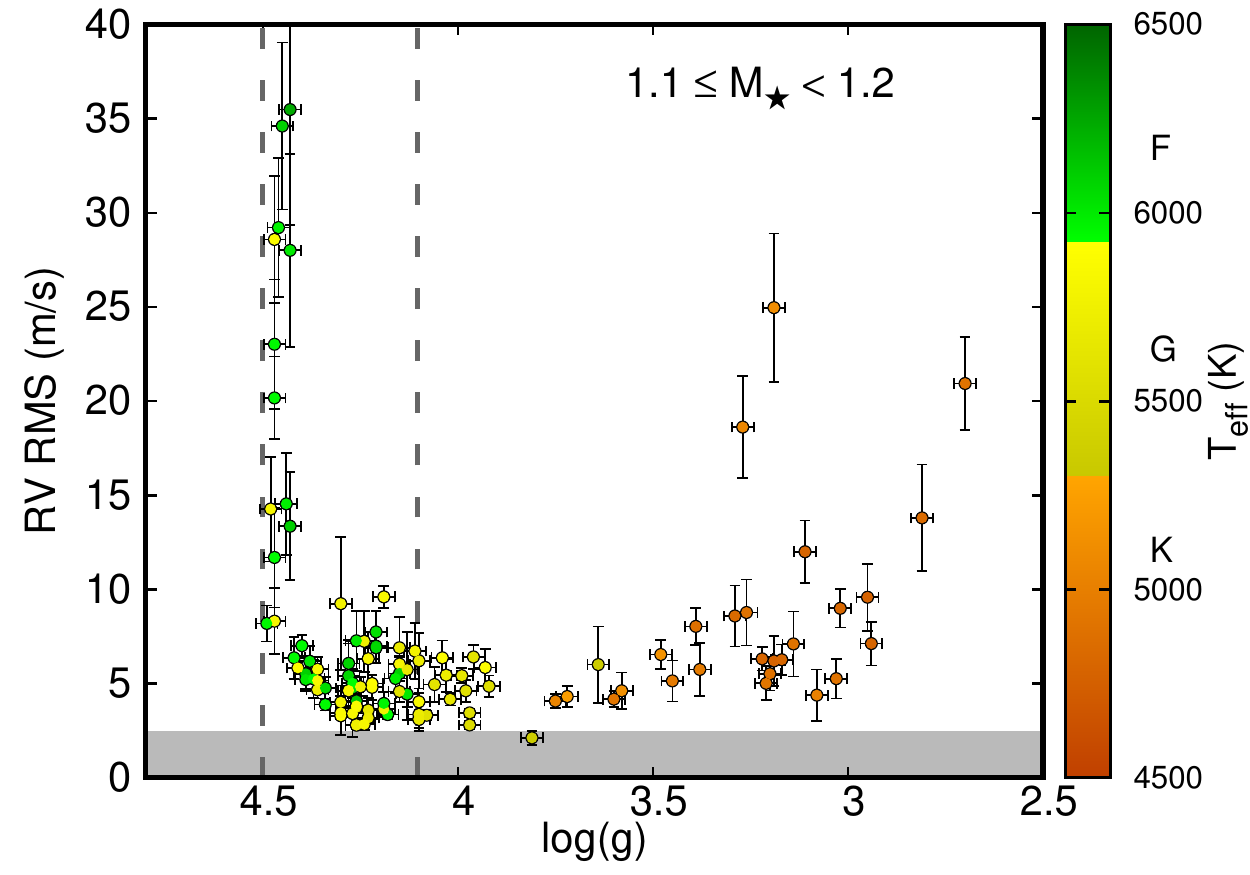}
\includegraphics[width=0.415\paperwidth]{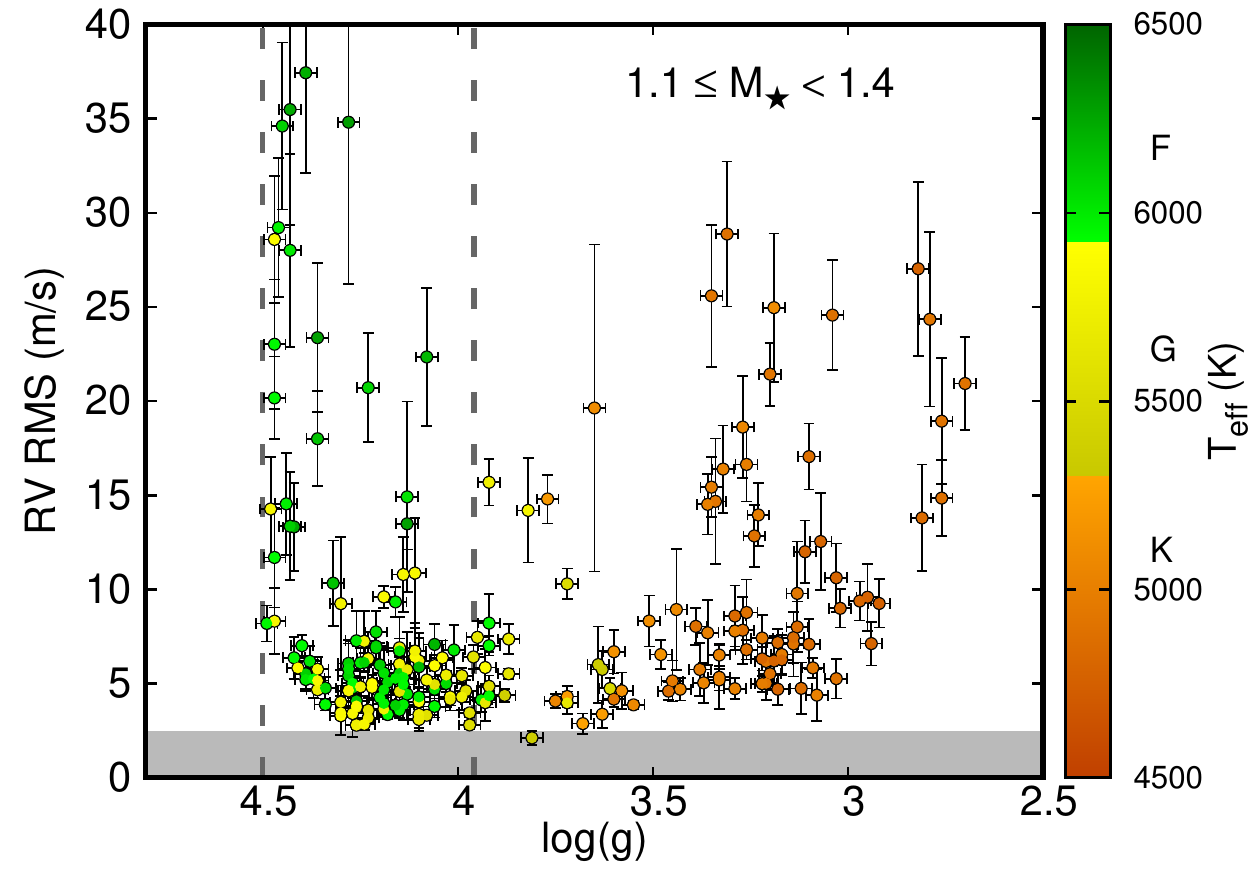}
\caption{RV jitter of Main Sequence and Retired (MSRF) stars as a function of surface gravity (reproduced from L20). The left figure, showing a narrow range of masses, highlights one such evolutionary track. The right figure shows the full sample of MSRF stars from L20. The vertical dashed lines in each figure indicate the locations of the zero age main sequence (ZAMS) and the terminal age main sequence (TAMS). Points are color-coded based on effective temperature to highlight the fact that main sequence stars show a gradient in temperature as they evolve \emph{across the main sequence} from ZAMS to TAMS. This further highlights why \emph{mass} is the most relevant quantity. This paper therefore refers to the mass range in the right panel as MSRF stars to clearly distinguish from purely spectral type ``F" star classification.}
\label{fig:f_star_jitter}
\end{figure*}

The focus of this work is to distinguish low jitter F stars from those with high jitter. Typical RV surveys target stars on the main sequence and as a result can use spectral type as a proxy for mass. However, spectral type classification as a proxy for mass based purely on temperature breaks down when considering evolved stars or even stars along the terminal age main sequence (TAMS), and so the quantity most relevant to planet-hunters is not the spectral type of the star, but the mass itself. Therefore we define our sample of `F' stars as stars in the mass range 1.1 to 1.4 M$_{\odot}$. These are stars that had spectral type F at their zero age main sequence (ZAMS). As shown in L20, clear relations between stellar properties and the radial velocity jitter arise when binning stars by mass. \autoref{fig:f_star_jitter} is a reproduction from L20 and shows the RV jitter as a function of the surface gravity for the mass bins of interest for this work. The stars are color-coded by spectral type ($\Teff$) to emphasize there is a clear temperature gradient even along the main sequence (between ZAMS and TAMS) and highlighting our reason to favor a mass-based definition over spectral type.

Since many of the stars in this mass range in our sample have evolved into G or K subgiants, we follow \citet{Johnson2006} and refer to stars in this mass range (1.1 to 1.4 M$_{\odot}$) as Main Sequence and ``Retired" F (MSRF) stars.

\subsection{Activity Metric}
We wish to look at the relation between jitter and chromospheric activity as measured by the emission peaks in the \ion{Ca}{2} H \& K lines. Two typical measurements of the chromospheric emission are \SHK{} and $\logRHK$. Both of these have historical basis in the Mount Wilson survey, which monitored the activity of hundreds of stars over several decades and was continued as a part of the California Planet Search program. The activity measurements in this analysis come from the latter, given that the California Planet search has the luxury of a spectral range that is useful for both Doppler measurements and contains the \ion{Ca}{2} H \& K lines at 3969 \AA{} and 3934 \AA, respectively. 

We choose to examine activity correlations with jitter by using the $\logRHK$ metric. This differs from \SHK, the main activity metric used in L20, in that $\logRHK$ subtracts out the photospheric component of the emission peak and also accounts for the star's color. This makes $\logRHK$ a better activity metric that can be compared between spectral types. We use an updated calibration from \citep{LorenzoOliveira2018} to derive $\logRHK$ using $\Teff$ rather than B-V. The usual caveat is that $\logRHK$ has not been calibrated for evolved stars, so although we derive $\logRHK$ values for these stars, the physical interpretation of the quantity as the fraction of the star's luminosity emitted in these chromospheric emission lines may not apply for the giants and subgiants.

\section{The Jitter Metric}\label{sec:jitter_metric}
We seek a way to reliably separate the high jitter MSRF stars from the low jitter MSRF stars. Based on \autoref{fig:f_star_jitter}, we see that the low jitter MSRF stars are those that are slightly evolved, either very near the TAMS or slightly more evolved than the TAMS. If one has precise mass and surface gravity measurements, one could easily tell if a given star falls in the expected range of $\logg$ where the jitter is below $\sim$10 m/s. However, precise masses and surface gravities are often expensive to obtain and are not available for many stars, so we seek a way to use more easily derived quantities. We instead restrict ourselves to $\Teff$, $\log{L}$, and $\logRHK$. 

We next construct a jitter metric based on the following intuition: we have seen that jitter decreases as the activity decreases before eventually increasing again due to convection (L20). If we also consider that stars below $\sim1.4$ M$_{\odot}$ show an increase in luminosity with evolution (both main sequence and post main sequence), then $\log{L}$ becomes a reasonable proxy for $\logg$ and we expect that stars should show an increase in jitter with increasing luminosity, also from convection. However, the stars with lowest $\log{L}$ are the youngest main sequence stars that are the most active and jittery. Therefore, the high jitter stars will have either a high value of $\logRHK$ or a high value of $\log{L}$. We therefore expect a simple function of the form $j = \alpha(\logRHK) + \log{L}$ should show a positive correlation with jitter, where $\alpha$ is chosen so that the sharp increase in jitter with $\logRHK$ (activity) closely matches the increase in jitter with $\log{L}$ (evolution).

\begin{figure}
\includegraphics[width=\columnwidth]{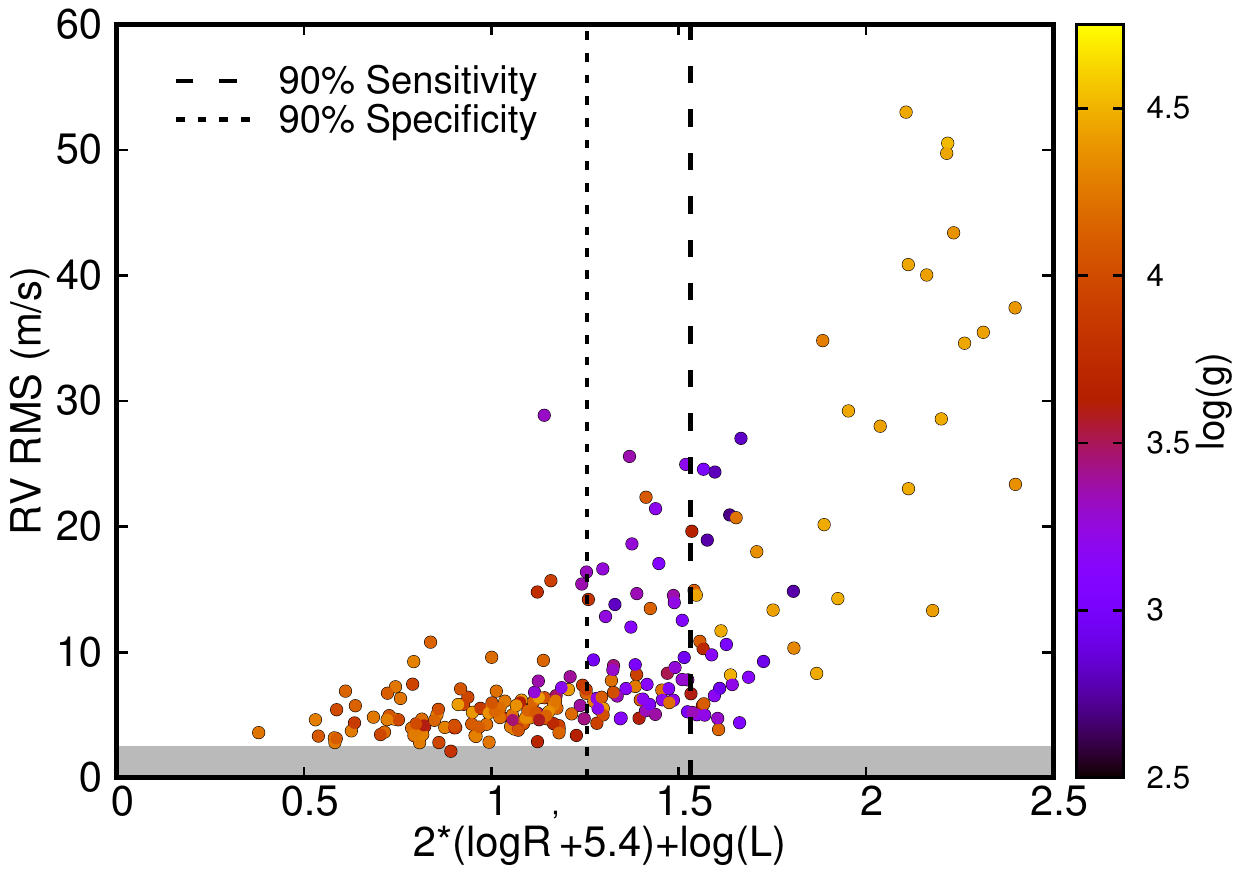}
\caption{RV jitter of  MSRF stars as a function of jitter metric $j$ (\autoref{eqn:jitter_metric}). Points are color-coded by surface gravity for clarity. The high jitter stars ($>10$ m/s) are generally either 1) subgiants with jitter dominated by convection or 2) active main sequence stars with jitter dominated by activity. The vertical line shows the chosen $j$ cut-off ($j_{0}=1.53$), chosen to have high sensitivity as described in \autoref{sec:low_jitter_f_stars}. The gray lines show the smoothed sensitivity, specificity, and PPV for the sample as a function of $j_{0}$ cutoff.}
\label{fig:jitter_metric_f}
\end{figure}

A simple linear regression recovers a best-fit value of $\alpha = 1.2$. However, we find that this value, despite a strong linear correlation between the RV jitter and the jitter metric $j$, does not do the best job of \emph{separating} the high jitter stars from the low jitter stars. We therefore choose $\alpha=2$ by eye after testing many values of $\alpha$ between 0 and 4. We adopt the jitter metric
\begin{equation}
j=2\left(\logRHK + 5.4\right) + \log{L}
\label{eqn:jitter_metric}
\end{equation}
where we have applied an offset of 5.4 to $\logRHK$ values to work with positive values. \autoref{fig:jitter_metric_f} shows a plot of RV jitter as a function of the jitter metric $j$ color coded by surface gravity. It is clear that the high jitter stars are either evolved stars with jitter dominated by convection or active stars with activity-dominated jitter as described in L20.

\section{Selecting Low Jitter F stars}\label{sec:low_jitter_f_stars}
We wish to suggest a cutoff in $j$ that is effective at selecting the low jitter ($<10$~m/s) MSRF stars. The exact $j$ threshold one uses depends on the motivations for selecting low jitter MSRF stars. For instance, one might desire a large sample of low jitter stars and wish to identify all of the low jitter stars for a big campaign. For this case they want a very low false negative rate to ensure they find all of the low-jitter stars in the sample.  Such a sample would be selected for ``completeness". On the other hand, one might be limited by telescope time and want to find a small number of guaranteed low-jitter stars to do followup on, in which case they need a very low false positive rate so that they don't waste any time at all on high jitter stars. This sample would be selected for ``pureness".

When implementing a cutoff in $j$, it is therefore useful to examine the \emph{sensitivity}, \emph{specificity} and the \emph{positive predictive value} of that cutoff to diagnose how effective it is at distinguishing low jitter stars in the two scenarios described above. These diagnostics are reviewed below.

\subsection{Sensitivity, Specificity, and Positive Predictive Value}
For each of these diagnostics, we split the sample two ways: first into ``true" and ``false" based on if they are low jitter ($<10$~m/s), and then into ``selected" and ``unselected" based on the $j$ threshold (``selected" stars are those below the threshold $j_{0}$). 

\begin{figure}
\includegraphics[width=\columnwidth]{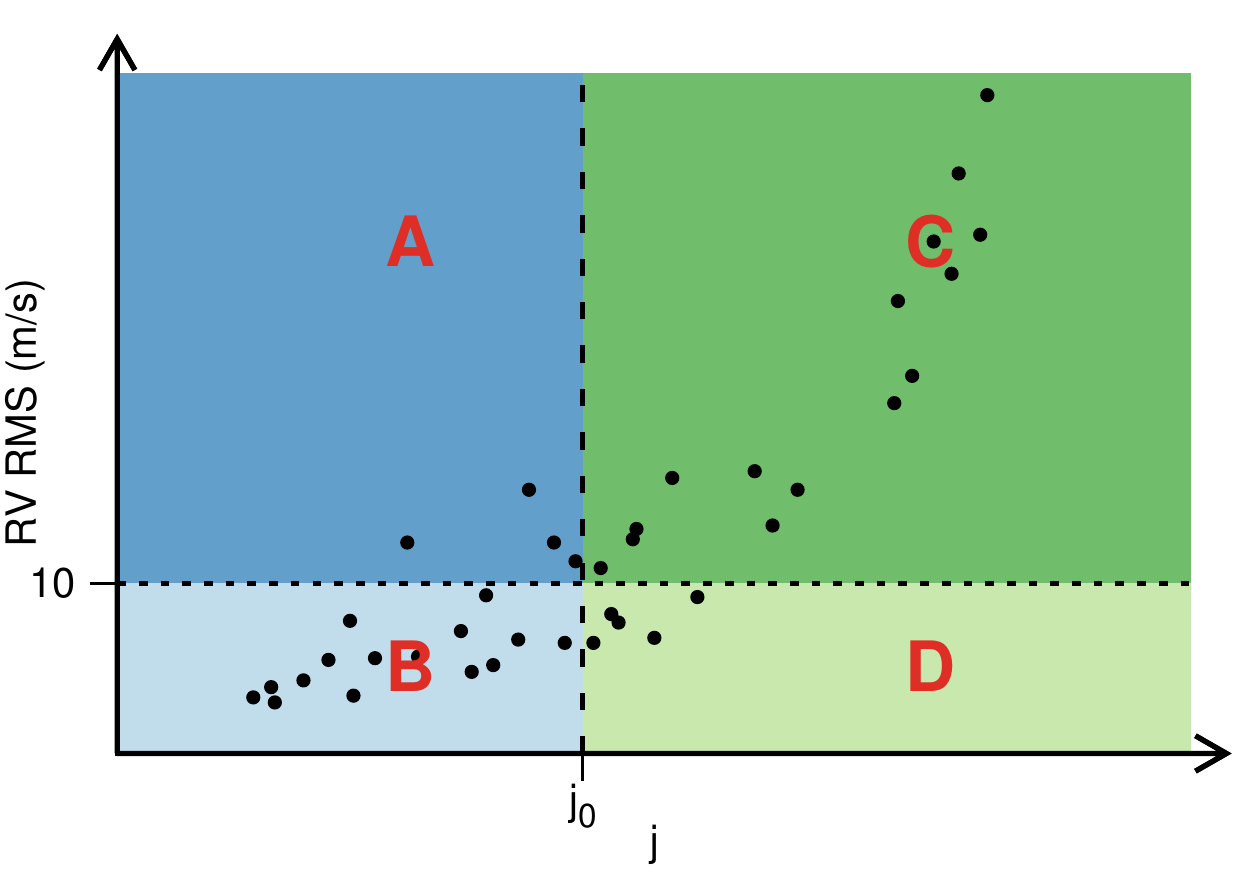}
\caption{Diagram (not actual data) showing regions useful for diagnosing the effectiveness of selecting low jitter stars. Using a $j$ threshold at $j_{0}$ (vertical dashed line) means that the blue regions indicate the ``selected" stars of a given sample, and the green regions indicate the rejected stars. The 10 m/s low jitter cutoff is denoted by the horizontal dashed line, separating the low jitter stars (``true" stars, light colored regions) from the high jitter stars (``false" stars, dark colored regions). Sensitivity is defined as B/(B+D), Specificity is C/(C+A), and PPV is B/(B+A).}
\label{fig:diagnostic}
\end{figure}

\paragraph{Sensitivity}
For a given $j$ threshold, $j_{0}$, the \emph{sensitivity} of the selection method is equivalent to asking: ``If a star has low jitter, how often will $j_{\star} < j_{0}$?". The sensitivity is defined as the number of stars that are selected and true (true positives) divided by the total number of ``true" stars. This can be seen graphically in \autoref{fig:diagnostic} as taking B/(B+D). Selecting $j_{0}$ for high \emph{sensitivity} therefore is useful for collecting a sample with as many low jitter stars as possible, a ``complete" sample.

\paragraph{Specificity}
Specificity is the complement to sensitivity. For a given $j$ threshold, $j_{0}$, the \emph{specificity} of the selection method is equivalent to asking: ``If a star has high jitter, how often will $j_{\star} > j_{0}$?". The specificity is defined as the number of stars that are rejected and false (true negatives) divided  by the total number of ``false" stars. This can be seen in \autoref{fig:diagnostic} as taking C/(C+A). Selecting $j_{0}$ for high \emph{specificity} is therefore useful for collecting a sample that excludes high jitter stars. By excluding the high jitter stars, selecting $j_{0}$ for high \emph{specificity} is therefore the better approach to selecting a ``pure" sample.

\paragraph{Positive Predictive Value}
For a given $j$ threshold, $j_{0}$, the \emph{positive predictive value} (PPV) of the selection method is equivalent to asking: ``If $j_{\star} < j_{0}$ what is probability that the star is low jitter?". PPV is defined as the number of stars that are selected and true (true positives) divided by the total number of selected stars. This can be seen in \autoref{fig:diagnostic} as taking B/(B+A). PPV is not monotonic with $j_{0}$ as the denominator depends on how many stars are below $j_{0}$ and for this reason, it is a more useful diagnostic once one has selected $j_{0}$ for either high sensitivity (``complete") or high specificity (``pure").

\subsection{Selecting the $j_{0}=1.53$ Threshold}
Our purpose was to establish two thresholds in $j_{0}$ to recommend in order to choose a ``complete" or ``pure" sample. In truth, PPV is a more intuitive diagnostic for the ``purity" of the sample than specificity is. However, because it is not monotonic with $j_{0}$ and is not normalized to be between 0 and 1 (the lower limit will always be determined by the percentage of low jitter stars in the entire sample), PPV does not make a useful diagnostic for selecting a threshold, but rather for diagnosing the effectiveness of a given threshold at selecting a ``pure" sample. In practice we found that choosing a threshold for high sensitivity resulted in high PPV as well. For that reason, we have opted to simply choose a threshold based solely on sensitivity. We show in \autoref{fig:jitter_metric_f} how sensitivity, specificity, and PPV vary with $j_{0}$ to allow the option for one to choose their own value of $j_{0}$ that suits their own unique purposes. These curves have been smoothed using a boxcar smoothing with width 0.15.

For the sample of MSRF stars in \autoref{fig:jitter_metric_f}, we choose the value of $j$ that produces 90\% sensitivity. This value is $j_{0}=1.53$, which is shown as the solid black line in \autoref{fig:jitter_metric_f}. This is the value of $j_{0}$ that selects 90\% of all low jitter MSRF stars. For this value of $j_{0}$, the specificity is 62\%. The PPV for this threshold is 86\%, meaning that among our sample of MSRF stars, this threshold results in 86\% low jitter stars.

We note that we have used 10 m/s as the threshold for low jitter. In fact, many of these stars have jitter below 7 and even 5 m/s. In general, there is a positive correlation between $j$ and RV RMS such that a lower $j$ value corresponds to a lower RV RMS.

\section{Selecting Low Jitter MSRF Stars Using Restricted Observables}\label{sec:restricted}
The previous section described a threshold in the jitter metric $j$ (\autoref{eqn:jitter_metric}) that can be used to select low jitter MSRF stars. However, as described in \autoref{sec:f_stars}, we defined as MSRF stars as those with masses between 1.1 and 1.4 M$_{\odot}$. We have therefore described a method to identify low jitter stars among 1.1 to 1.4 M$_{\odot}$ stars. As mentioned previously, we will not always have precise masses for stars \emph{a priori}, and so the remainder of this work is focused on taking the lessons learned by working in terms of mass and transforming into more readily-available quantities. 

\begin{figure*}
\centering
\includegraphics[width=0.415\paperwidth]{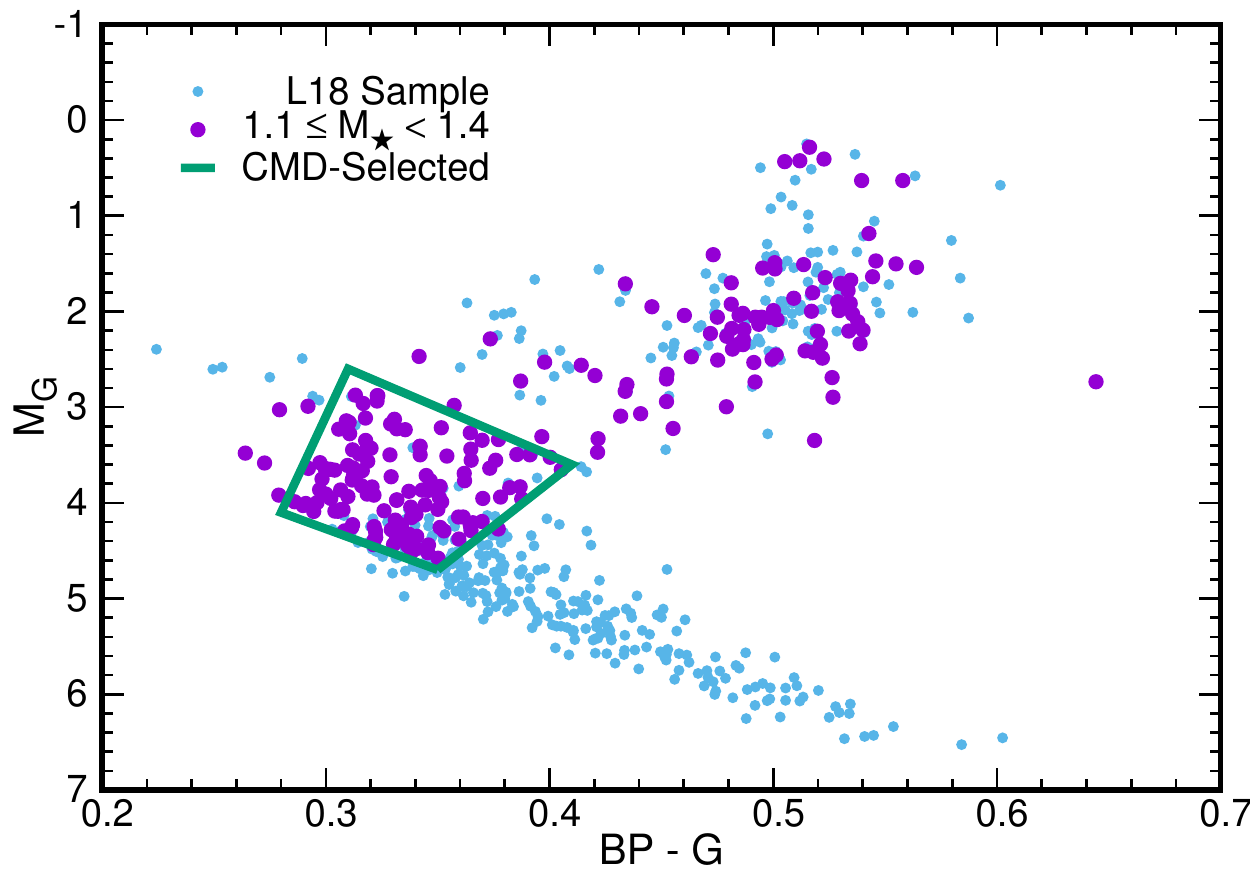}
\includegraphics[width=0.415\paperwidth]{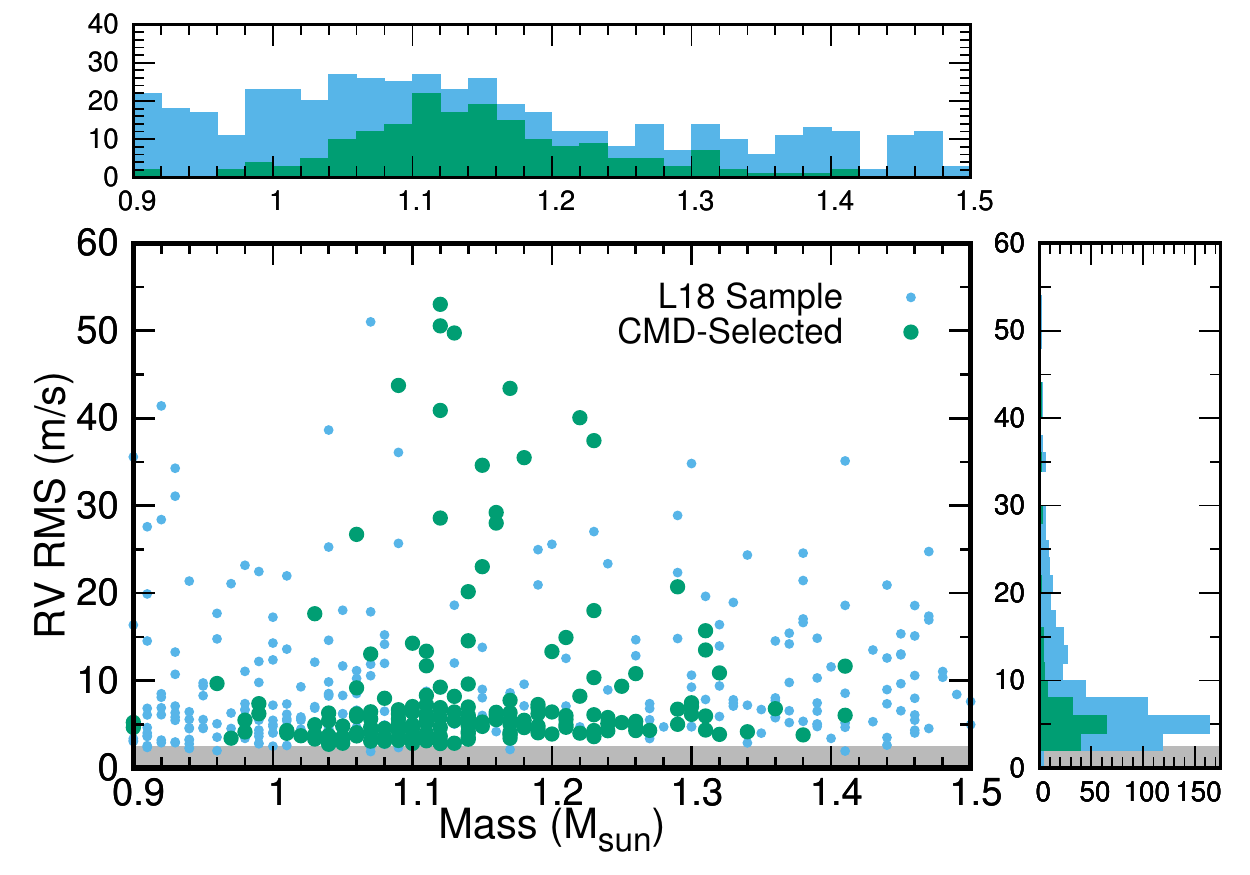}
\caption{(\emph{Left}) Gaia color-magnitude diagram of the L20 sample (light blue) with the stars in the 1.1 to 1.4 M$_{\odot}$ mass bin (MSRF stars) shown in purple. The box shows our selection region to best re-select these stars without prior knowledge of their mass. We have purposefully not included the most evolved subgiant and giant stars to avoid contamination from stars with higher masses. The excluded subgiants are likely to be stars with jitter dominated by convection (L20), only a handful of which will have low jitter. The selection box is drawn from the vertices (0.35,4.7), (0.41,3.6), (0.31,2.6), (0.28,4.1). (\emph{Right}) RV jitter of the two samples of stars as a function of mass. The selected region does a sufficient job of selecting stars in the mass range 1.1 to 1.4 M$_{\odot}$, although there are a number of stars below 1.1 M$_{\odot}$ (seen in the mass histogram above) with this method. We note that the CMD selection is used purely to select MSRF stars, not necessarily \emph{low jitter} MSRF stars.}
\label{fig:gaia_CMD_selection}
\end{figure*}

The first step is to start with the entire sample from L20 and attempt to re-select the MSRF mass bin using the Gaia color-magnitude diagram (CMD). \autoref{fig:gaia_CMD_selection} shows the CMD diagram for the L20 sample with stars in the 1.1 to 1.4 M$_{\odot}$ mass bin shown in purple. We have drawn a box by eye in the CMD to best select those stars using a polygon with vertices (0.35,4.7), (0.41,3.6), (0.31,2.6), (0.28,4.1). We note that we have purposefully removed most evolved subgiant/giant stars in this sample because stars of a wide range of masses occupy the same region of the CMD. More importantly, the subgiant and giant stars generally have high jitter due to convective motions (see \autoref{fig:f_star_jitter}). However, from \autoref{fig:jitter_metric_f} we see that these stars (in purple) will contain a mix of both high jitter and low jitter stars in the range $1.3 < j < 1.7$. We emphasize that this selection cut in the Gaia CMD is chosen only to best select MSRF stars, not necessarily to select \emph{low jitter} MSRF stars. 

The mass histogram in the right panel of \autoref{fig:gaia_CMD_selection} shows how successful the Gaia CMD selection cuts are at selecting stars between 1.1 and 1.4 M$_{\odot}$. While this clearly also includes stars outside of this mass range the majority of those stars are only barely outside the range toward the lower mass end, in the range 1.0 to 1.1 M$_{\odot}$. At 25\%, the large fraction of stars that are slightly more massive than solar (1.0 to 1.1 M$_{\odot}$) is not ideal for reproducing the MSRF star sample, but these are likely early G stars that are similarly avoided in RV surveys for their expected levels of jitter. As such, their inclusion will be useful for showing the utility of the jitter metric $j$ at selecting low jitter stars. The missing stars in the 1.1 to 1.4 M$_{\odot}$ range are the evolved subgiant and giant stars, as described above.

Now that we have selected a sample that contains mostly MSRF stars, we can investigate how well the jitter metric from \autoref{sec:low_jitter_f_stars} that we used in MSRF-only sample apply. We go one step further and define a second jitter metric where the luminosity term is replaced by the Gaia absolute G magnitude. This second jitter metric, which we call $j'$, is defined
\begin{equation}
j'= 2(\logRHK + 5.4) + 0.5(4.5-{G})
\label{eqn:j_prime}
\end{equation}

\begin{figure*}
\centering
\includegraphics[width=0.415\paperwidth]{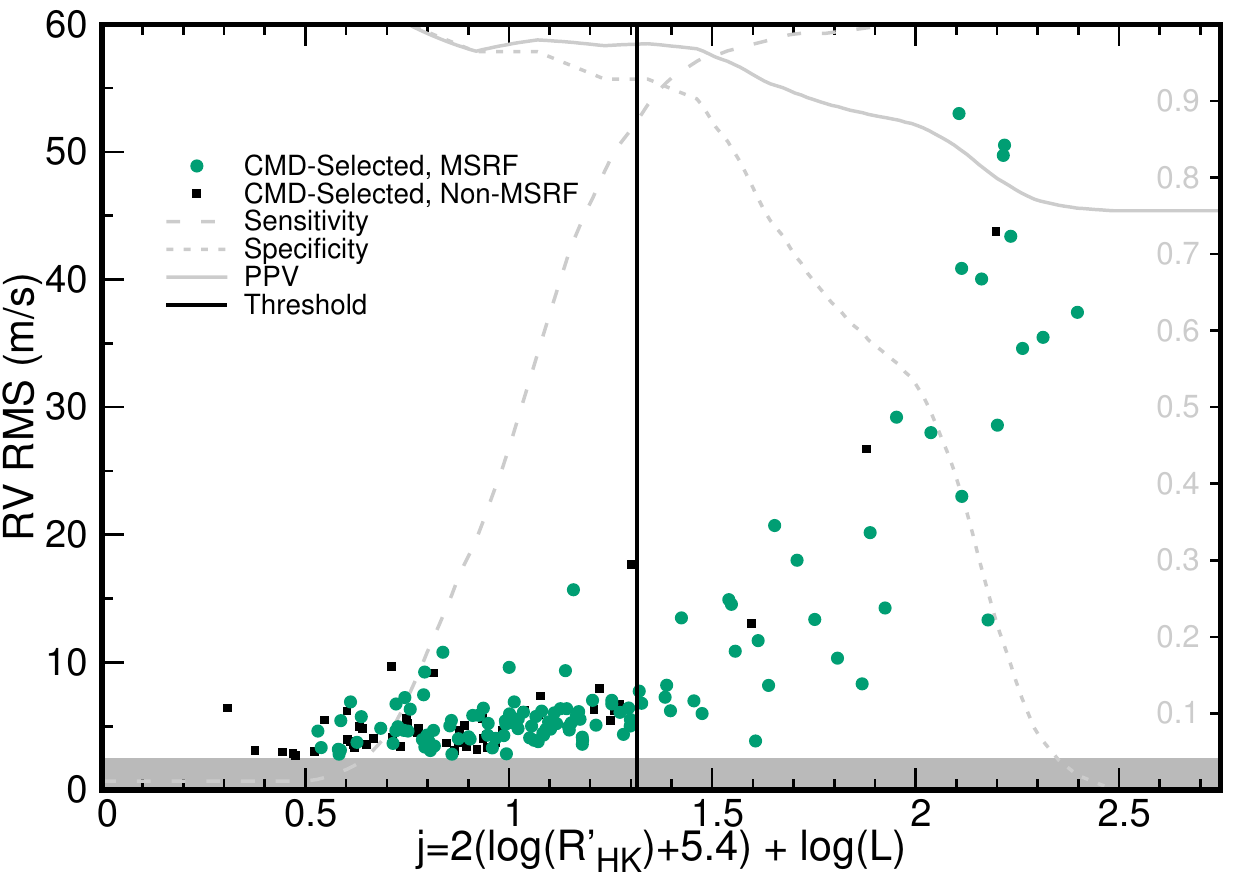}
\includegraphics[width=0.415\paperwidth]{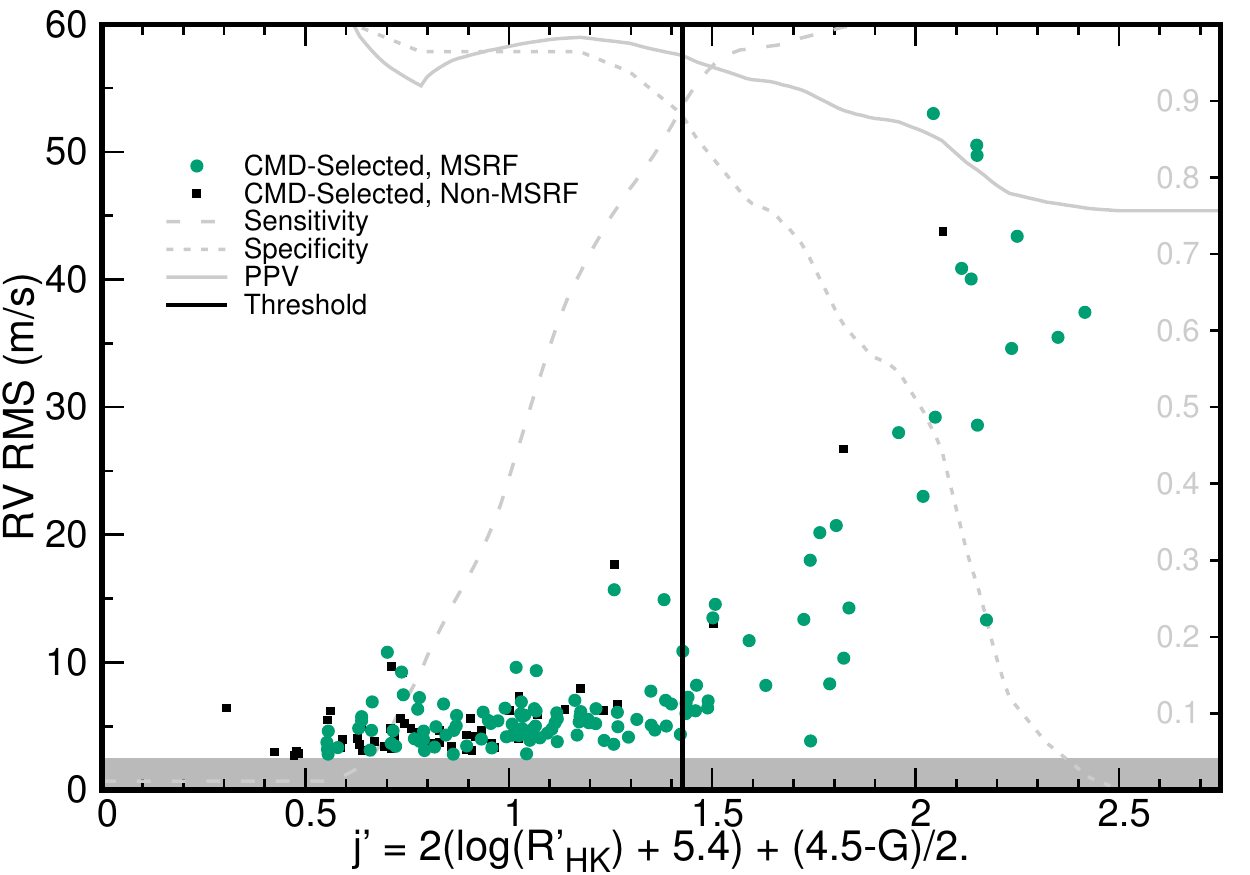}
\caption{(\emph{Left}) Jitter metric $j$ for the stars in L20 that lie with the region shown in \autoref{fig:gaia_CMD_selection}. The MSRF stars within this region are shown in green, with the non-MSRF stars shown in black. The vertical line shows the jitter metric threshold to best select low jitter stars based on 90\% sensitivity ($j_{0}=1.32$). The gray lines show the smoothed sensitivity, specificity, and PPV curves as a function of $j_{0}$. (\emph{Right}) A second jitter metric which uses absolute G magnitude instead of luminosities. This second jitter metric, which replaces the luminosity term with a Gaia G magnitude term, is nearly identical to the first. The vertical line again shows the 90\% sensitivity ($j'_{0}=1.43$). In both figures, the $j_{0}$ and $j'_{0}$ thresholds have been selected using only the MSRF stars (green points) within the CMD-selected region. See \autoref{sec:restricted} for details on how the sensitivity, specificity, and PPV change when including the non-MSRF stars.}
\label{fig:jitter_metric_selection}
\end{figure*}

We note that $j$ and $j'$ are very well correlated as expected and show linear agreement with standard deviation of 0.15; the only reason that there is not an exact linear agreement is because the luminosities reported in L20 (and used in $j$) are those of \citet{Brewer2017}, who used \emph{Hipparcos} distances. \autoref{fig:jitter_metric_selection} shows the RV RMS of the CMD-selected stars as a function of the jitter metric $j$ and as a function of $j'$. Following our previous example, we again choose thresholds in $j$ and $j'$ that correspond to 90\% sensitivity. To select these thresholds, we use only the stars in the CMD-selected sample that are in fact within the MSRF mass range. We do this to avoid influence from the non-MSRF stars that are generally below the mass range, with lower $j$ and lower jitter. The 90\% sensitivity threshold among CMD-selected MSRF stars is $j_{0}=1.32$, which results in 93\% specificity, and 98\% PPV. When we include the entire sample of CMD-selected stars, this $j_{0}$ corresponds to 93\% sensitivity, 91\% specificity, and 98\% PPV. 
We repeat this exercise for $j'$. The 90\% sensitivity threshold among CMD-selected MSRF stars is $j'=1.43$, which results in 86\% specificity and 96\% PPV. When we include all CMD-selected stars, this $j'_{0}$ threshold corresponds to 93\% sensitivity, 85\% specificity and 96\% PPV.

\section{Selecting Low Jitter MSRF Stars Using Gaia Only}\label{sec:gaia_only}
In the previous sections we have identified low jitter MSRF stars first by using two different methods of selecting MSRF stars (mass-based as defined, and using the Gaia CMD) and then using two different ``jitter metrics" ($j$ and $j'$) to select the low jitter stars of those samples. The jitter metrics used to distinguish the low jitter stars from the high jitter stars have both required an activity measurement, $\logRHK$, which for many stars may not be available. The purpose of this section is to try to remove the activity dependence and describe a best practice for selecting low jitter stars using data only from Gaia DR2 \citep{GaiaDR2}, which is readily available for millions of stars.

\begin{figure}
\includegraphics[width=\columnwidth]{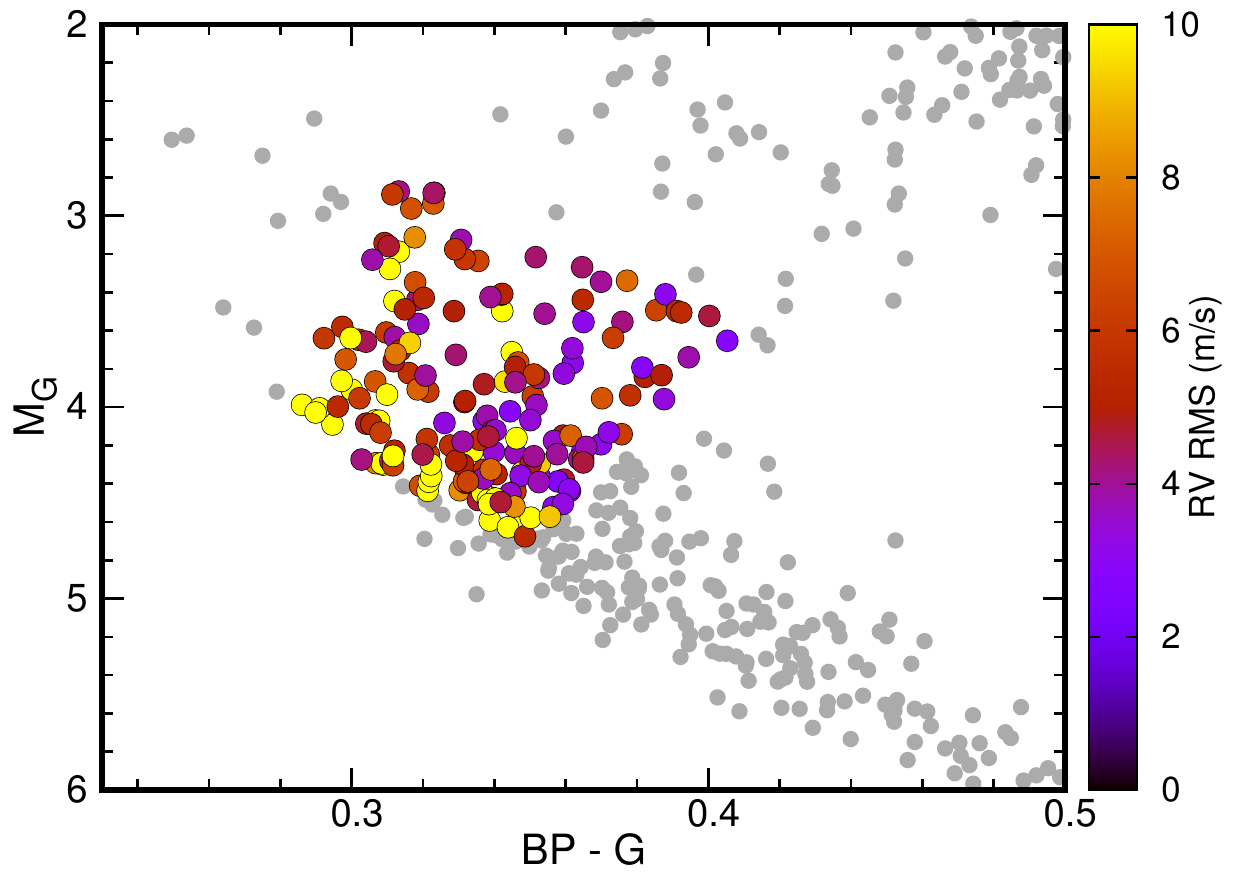}
\caption{Gaia CMD highlighting the jitter gradient of the CMD-selected stars perpendicular to the main sequence. Yellow points are those with RV jitter greater than 10 m/s. The gray points show the rest of the L20 sample not in the CMD-selected region. The drop in jitter with evolution off of the main sequence is clear, however the more luminous evolved stars have somewhat higher jitter because they are still rapidly rotating and active, which is captured by the activity term in the $j$ metric.}
\label{fig:gaia_cmd_jitter}
\end{figure}

Our intuition is as follows: we expect from \autoref{fig:f_star_jitter} that the highest jitter MSRF stars are the active stars nearest the ZAMS and the lowest jitter MSRF stars are those nearest to the TAMS and should therefore expect to see a jitter gradient in the main sequence of an HR diagram that is perpendicular to the main sequence. We show this gradient in the Gaia CMD in \autoref{fig:gaia_cmd_jitter}. With this knowledge, there should therefore be a small region in the HR diagram where we expect the majority of low-jitter MSRF stars to reside. 

Of course, this perpendicular gradient is tied to the surface gravity and so we will use this fact to help select the low jitter MSRF stars. Rather than the surface gravity, for which precise measurements are not available for most stars, we seek a Gaia-based evolution metric similar to that of \citet{Wright2004b}.

\subsection{Gaia Main Sequence and Evolution Metric}
\citet{Wright2004b} used HIPPARCOS data to fit a high-order polynomial to the main sequence and then calculated the height of stars in V-band magnitude above the main sequence to estimate the evolution of the star. Here we present a similar calculation using the Gaia DR2 data.

\begin{figure}
\includegraphics[width=\columnwidth]{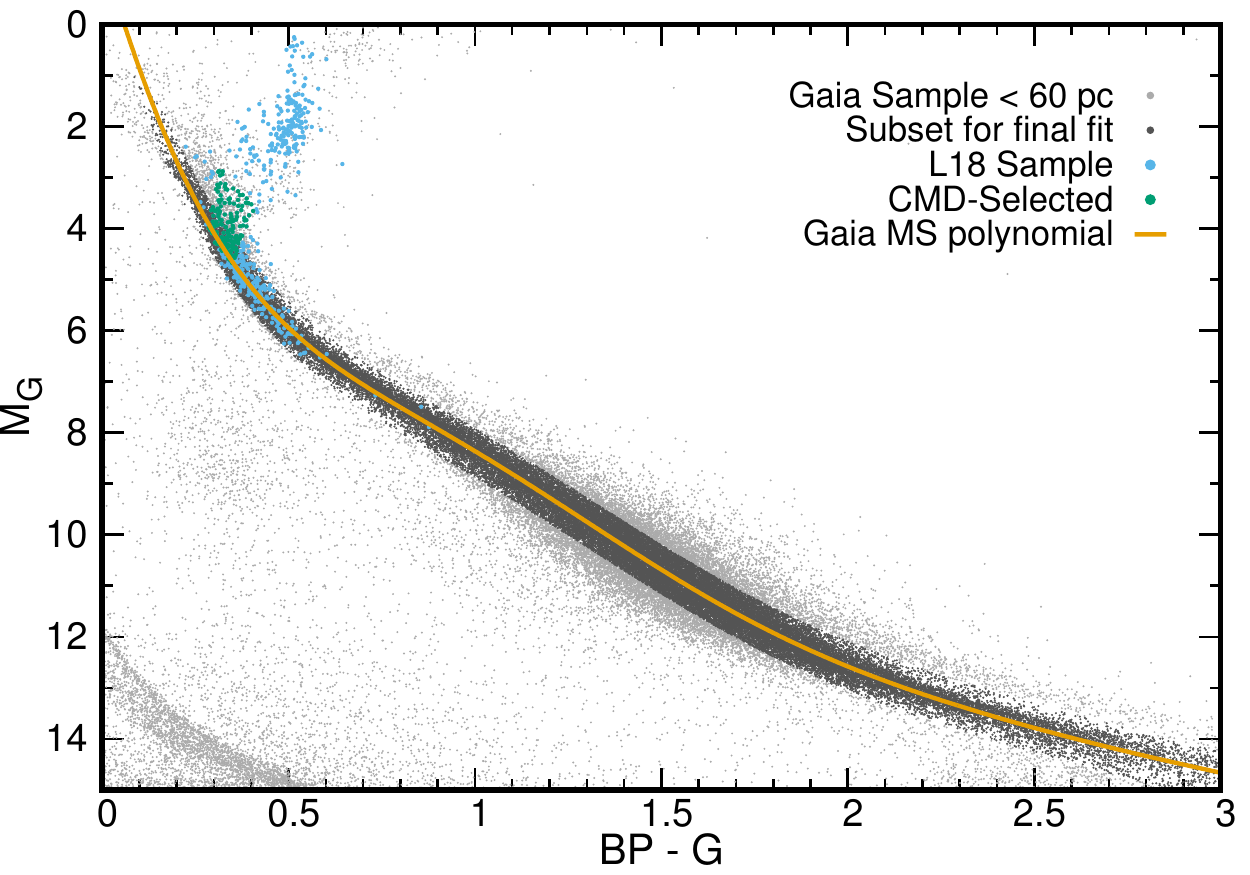}
\caption{The Gaia DR2 main sequence. Light gray points show the Gaia DR2 sample within 60 pc that were used to fit iteratively higher-order polynomials to the main sequence. The dark gray points show the points remaining in the final iteration used to fit the 9th-order polynomial shown in the black line. The L20 sample (blue) and CMD-selected stars (green, see \autoref{fig:gaia_CMD_selection}) are plotted for reference.}
\label{fig:gaia_ms}
\end{figure}

First, we selected the Gaia stars within 60 pc and constructed a Gaia CMD again using Gaia G band absolute magnitude and using the G$_{BP}$-G color, where G$_{BP}$ is the Gaia blue passband \citep{Riello2018}. We started with a by-eye linear `fit' to the data, and selected all stars within 3 G magnitudes of the linear fit. We then iteratively fit higher order polynomials, each time fitting the stars within 2$\sigma$ of the fit. In the end we get a 9th order polynomial of the Gaia Main Sequence such that 
\begin{equation}
G_{\mathrm{MS}} = \sum_{n=0}^{9} a_{n}(G_{BP}-G)^n,
\end{equation}
where $a=$\{-1.4061931, 24.892279, -21.260701, -12.473151, 44.584722, -40.649264, 18.977568, -4.9528572, 0.68840473, -0.039803364\}. The Gaia stars and main sequence polynomial are shown in \autoref{fig:gaia_ms}. This polynomial is applicable over the range $0 < \mathrm{BP} -\mathrm{G} < 3$.

The height above the main sequence, $\Delta G$, is therefore
\begin{equation}
\Delta {G} = G_{\mathrm{MS}}(G_{BP}-G)-G.
\label{eqn:delta_mg}
\end{equation}

\begin{figure}
\includegraphics[width=\columnwidth]{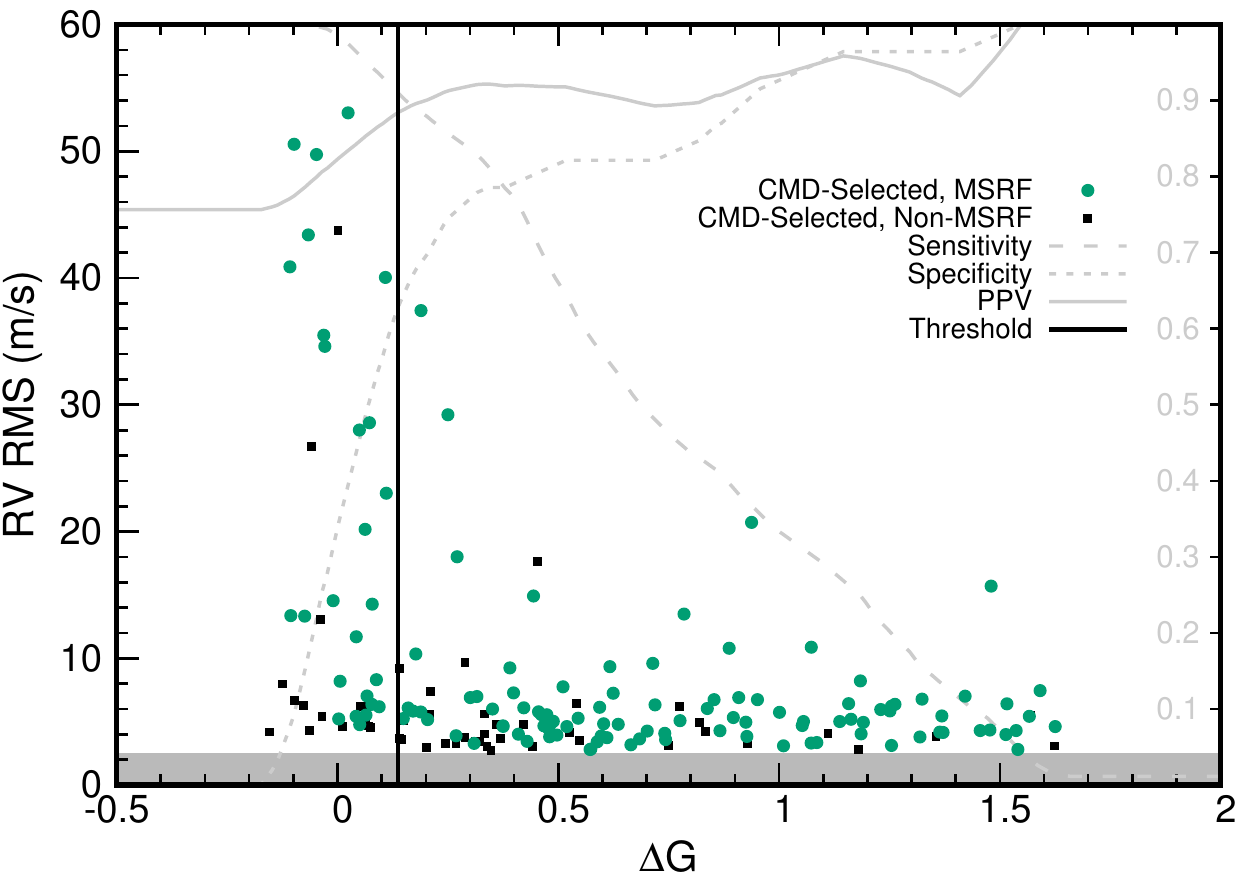}
\caption{RV jitter of CMD-selected MSRF stars as a function of height above the main sequence, $\Delta G$ (\autoref{eqn:delta_mg}), a proxy for evolution. By selecting the MSRF stars with $\Delta G > 0.14 $, we can achieve sensitivity of 90\%. Gray lines show the smoothed sensitivity, specificity, and PPV for the sample as a function of $\DeltaMG$ threshold.}
\label{fig:gaia_only}
\end{figure}

By plotting RV jitter as a function of $\DeltaMG$ \autoref{fig:gaia_only}, we see that we can employ the same diagnostics used earlier with the jitter metric in order to select for high sensitivity by using $\DeltaMG$ in place of $j$ or $j'$. We again wish to distinguish between the actual MSRF stars in the CMD-selected region and the non-MSRF stars in that region. 

The 90\% sensitivity threshold among CMD-selected MSRF stars is $\DeltaMG=0.14$, which results in a 65\% specificity and 90\% PPV. When including the non-MSRF stars, this value of $\DeltaMG$ corresponds to 86\% sensitivity, 65\% specificity, and 91\% PPV. 

To summarize, the height above the main sequence is an imperfect measure of jitter alone, but combined with the selection cut in the CMD it provides a way to achieve 90\% sensitivity in selecting low jitter MSRF stars without prior knowledge of $\logRHK$ or mass.

\section{Discussion}\label{discussion}
We have shown in this work that we are able to select low jitter MSRF stars with a high degree of certainty. The PPV values for each of the suggested selection thresholds in this paper are quite high ($\sim$90\%), meaning that for any given selection threshold in this work, $\sim$90\% of the selected stars will be low jitter stars. One caveat of this is that when using the Gaia CMD to select the MSRF stars, not all stars in the sample will actually be MSRF stars.

It is also necessary to address the bias of this sample. We are working specifically with stars selected for RV surveys with measured jitter in L20. It is likely that this sample will be inherently biased toward low jitter stars. However, it is beyond the scope of this paper to estimate the degree of this bias. An additional factor is that L20 required 10 observations for a reliable jitter to be measured. This could also skew the sample toward lower jitter stars, as observations of many stars could have been aborted after several observations showed too high of jitter to be useful for planet-hunting. In general, we believe this to be a low-order effect as evidenced by the many high jitter stars with plenty of observations already present in the sample. It is most likely that a more representative sample of MSRF stars would contain many more high jitter stars, however, the statistical sample provided by L20 suggests that they would follow the same trends with $j$, $j'$, or $\DeltaMG$, and so this would have the most effect in specificity and PPV values for the thresholds beyond the high sensitivity thresholds presented here (for instance, $j >1.75$ in \autoref{fig:jitter_metric_f}).

\section{Conclusions and Summary}\label{sec:summary}
Using a subset of stars from L20 we have shown that stars of mass 1.1 to 1.4 M$_{\odot}$, which we refer to as Main Sequence and Retired F (MSRF) stars, can be RV stable to $< 5$ m/s. These stars have jitter that is dominated by convection rather than activity-dominated features. As such, using spot models or other activity-related methods of reducing jitter will be less effective for discovering low mass planets at or below this amplitude. Instead, these stars will have jitter dominated by granulation/oscillations and different observational techniques are necessary to beat down the jitter to tease out the signals of low mass planets.

The presence of RV stable MSRF stars calls into question past surveys that have specifically avoided them because of their high level of expected jitter. To better inform future RV observations, we developed a jitter metric (\autoref{eqn:jitter_metric}) that takes into account the competing age effects of RV jitter: activity-dominated jitter that decreases with age and convection-dominated jitter that increases with age. We have shown that this metric can be used to select low jitter MSRF stars using a high sensitivity threshold.

In an effort to be as useful to the exoplanet community as possible, we then restricted ourselves to readily-available observables, and selected a new sample of mostly MSRF stars based on a selection cut in the Gaia color-magnitue diagram. Applying the same jitter metric to these stars showed good agreement with the previous sample of MSRF stars and enabled selection of low jitter stars again with high sensitivity. To simplify it even more, we adopted a second jitter metric which uses only an activity measurement and the Gaia G-band magnitude, which performs similarly.

Finally, we restricted ourselves even more by removing the need for an activity metric. We use recent Gaia DR2 data to fit a polynomial to the Gaia main sequence and define a proxy for evolution based on a star's height above the main sequence. Using this metric and the selection cuts in the Gaia CMD, we are again able to select low jitter MSRF stars with both high sensitivity and high specificity.

To summarize:

\begin{enumerate}
\item \textbf{If precise masses and activity measurements are available}, a 90\% sensitivity threshold in $j$ (\autoref{eqn:jitter_metric}) of $j_{0}=1.53$ is effective at selecting low jitter MSRF stars, with a PPV of 86\%.
\item \textbf{If activity measurements are available but precise masses are unavailable}, the Gaia CMD selection cuts can be used to construct a sample of mostly MSRF stars. Then 90\% sensitivity threshold in the jitter metric $j$ (or alternative $j'$, see \autoref{sec:restricted}) of $j_{0}=1.32$ is effective at selecting low jitter MSRF stars, with PPV of 62\%. The lower PPV is driven by the performance of selecting purely MSRF stars. Note that if one is not concerned that the exact mass is in the MSRF range and cares more about the PPV of selecting a low jitter star of \emph{any} mass in this CMD selection, the PPV becomes 98\%.
\item \textbf{If activity measurements are unavailable and one wishes to use only Gaia data}, the jitter metric $j$ can no longer be used because of its dependence on activity ($\logRHK{}$). Instead, we suggest using $\DeltaMG$, the height above the main sequence, as the metric for selecting low jitter MSRF stars using $\DeltaMG=0.14$, again chosen for 90\% sensitivity. The PPV for this threshold is 64\%. Again, the low PPV value comes from the ability to select only MSRF stars using the Gaia CMD. Note that if one is not concerned that the exact mass is in the MSRF range and cares more about the PPV of selecting a low jitter star of \emph{any} mass in this CMD selection, the PPV becomes 91\%.
\end{enumerate}

The various methods for selecting low jitter MSRF stars in this paper will not only be useful for opening a currently-avoided region of planet searches, but also for determining the best candidates for RV followup from transit surveys such as \emph{K2}, \emph{TESS}, \emph{PLATO}, or \emph{CHEOPS}.

\acknowledgements{
The authors thank Fabienne Bastien for her many discussions that helped shape the focus of this work. We also thank Andrew Howard for various discussions regarding the California Planet Search and its continued observations.

We thank the anonymous referee for their speedy report, which helped clarify many of the points in this work.

The authors wish to recognize and acknowledge the very significant cultural role and reverence that the summit of Maunakea has always had within the indigenous Hawaiian community.  We are most fortunate to have the opportunity to conduct observations from this mountain.

Keck time for this project has been awarded from many sources, primarily institutional time from the University of California, Caltech, NASA, and Yale. We thank the many observers and CPS team members who have worked over the decades to produce this invaluable data set.

This research has made use of the SIMBAD database, operated at CDS, Strasbourg, France; the Exoplanet Orbit Database and the Exoplanet Data Explorer at exoplanets.org.; and of NASA's Astrophysics Data System Bibliographic Services. This work was partially supported by funding from the Center for Exoplanets and Habitable Worlds, which is supported by the Pennsylvania State University, the Eberly College of Science, and the Pennsylvania Space Grant Consortium. This material is based upon work supported by the National Science Foundation Graduate Research Fellowship Program under Grant No. DGE1255832.
}

\bibliography{\string~/Google_Drive/Research/library}

\end{document}